\documentclass[aps,prb,preprint,superscriptaddress]{revtex4}

\usepackage{amssymb} 

\usepackage{graphicx}
\usepackage{verbatim}
\usepackage{bbm}

\date{\today}

\newcommand{\SiH}[1]{Si$_{29}$H$_{24}$}
\newcommand{\Styr}[1]{C$_{8}$H$_{8}$}
\newcommand{\Styra}[1]{C$_{8}$H$^{A}_{8}$}
\newcommand{\Styrb}[1]{C$_{8}$H$^{B}_{8}$}

\begin{document}


\title{Negative differential resistance of Styrene on an ideal 
  Si[111] surface: dependence of the I-V characteristics on geometry, 
surface doping and shape of the STM-tip} 
\author{Samuel E. Baltazar} 
\email[]{sbaltaza@ipicyt.edu.mx}
\affiliation{Advanced Materials Department, IPICYT, Camino Presa San 
Jos\'e 2055, 78216, San Luis Potos\'{\i}, S.L.P., M\'exico}
\affiliation{Theoretische Physik, Fachbereich Naturwissenschaften, and 
Center for Interdisciplinary Nanostructure Science and Technology (CINSaT),  
Universit\"at  Kassel, Heinrich-Plett-Str. 40, 34132 Kassel, Germany.} 
\author{Mario De Menech}
\affiliation{Theoretische Physik, Fachbereich Naturwissenschaften, and 
Center for Interdisciplinary Nanostructure Science and Technology (CINSaT),  
Universit\"at  Kassel, Heinrich-Plett-Str. 40, 34132 Kassel, Germany.}
\affiliation{$^1$ Max--Planck--Institut f\"ur Physik komplexer Systeme\\
  N\"othnitzer Str.\,38, 01187 Dresden, Germany}
\author{Ulf Saalmann}
\affiliation{$^1$ Max--Planck--Institut f\"ur Physik komplexer Systeme\\
  N\"othnitzer Str.\,38, 01187 Dresden, Germany}
\author{Aldo H. Romero}
\affiliation{CINVESTAV Quer\'etaro, Libramiento Norponiente No 2000,  
76230 Quer\'etaro, M\'exico} 
\author{Martin E. Garcia} 
\affiliation{Theoretische Physik, Fachbereich Naturwissenschaften, and 
Center for Interdisciplinary Nanostructure Science and Technology (CINSaT),
 Universit\"at Kassel, Heinrich-Plett-Str. 40, 34132 Kassel, Germany.}

\begin{abstract} 
 We study the  electron transport properties through a supported
organic molecule styrene (\Styr~) on an ideal silicon surface Si[111] and probed by a STM-tip.
 The I-V characteristics and the differential conductance of the molecule 
are calculated using a self consistent approach based on non equilibrium 
Green's functions. 
 Two different adsorption configurations for the molecule on the surface 
 were considered which corresponds to a global and a local minimum of the total energy. 
 In both cases we  find a negative differential resistance (NDR) 
 in a given interval of bias voltages.
 This effect is controlled by the states available close to the 
Fermi level of the surface and can be manipulated by properly 
doping the substrate. 
 We also analyze the influence of the tip-shape on 
 the I-V characteristics. 
\end{abstract}


\maketitle

\section{Introduction}
Transport properties through supported molecules on substrates
are an interesting phenomena in the field of nanodispositives.
This idea has been
extensively studied in the last few years mainly due to the possibility to 
build electronic dispositives at atomic scale.
The major bottle neck in nanoelectronics is to have different mechanisms to control
the transport properties as wish. In order to get to this point 
experiments and theory should work together by investigating different
molecules, substrates and physical properties to achieve this goal.
In the case of substrates, two different approaches have been followed
by using metallic or semiconductor surfaces as the electrodes which
interact with a given molecule.
With metallic contacts, the conduction could be described
through an ohmic behavior, but if non-metallic contacts are considered
(where the surface can have gaps in the electronic dispersion relation
or if the molecule and surface
interaction is able to open a gap), the electronic transport could 
change. Changing the molecule type has been also further
investigated and many interesting results could be found in the literature.
Some examples about electronic transport in mesoscopic systems consider
carbon nanotubes used as nanowires ~\cite{Dekker1997}, metal-ion clusters useful
for information storage ~\cite{Sesoli1993} and
molecules such as small silicon clusters that
have been then studied showing resonant
tunneling effects due to the electronic states of the cluster~\cite{Bolotov2001}.
In the particular the case of silicon clusters in contact with semiconductor
surface have been theoretically investigated in our group. We did report
a diode-like behavior, opening the possibility of using a like dispositive in the
the design of new electronic dispositives~\cite{Baltazar2007}.
Channels responsible of the electronic
flow are not only established by the molecular levels of the molecule,
but also by the position of the Fermi level associated with the surface.
Recently, negative differential resistance (NDR) on heavily
doped Si substrates in contact with organic molecules
has been predicted theoretically~\cite{Rakshit2004} and demonstrated
experimentally~\cite{Guisinger2004}. This observation has been related to the
crossing of the HOMO or LUMO levels of the band edge of the underlying
semiconductor. It has been also pointed out that by using different molecules such as
metallic nanocrystals or organic molecules, the conductance can be changed to be
diode-like behavior and showing a dependence on the used molecules
~\cite{Reed1997,Klein1996}. These recent results have reinforced the idea
of using molecular systems as active structures  in electronic components
at nano scale.\\
In this work, we want to consider the electronic transport through an organic molecule
in contact with a
semiconductor surface. We report and explain a non-ohmic behavior in the electronic
transport and we relate this observation to the
interaction between the different systems. Transport properties are also tested under
different conditions, such as doping and applied external electric fields.

\section{Formalism}
In this section we draw the methodology we have followed to calculate the
transport properties here reported. This methodology has been previously
presented by De Menech to study silver clusters and C${}_{60}$ fullerenes 
on metal surfaces\textit{et al.} \cite{DeMenech2006a, DeMenech2006b}.
Our system consists of substrate-molecule-electrode. The surface has
been modeled as an ideal structure with no reconstruction after the
molecule has been absorbed on it. The electrode is a metallic tip far away
from the molecule, with no bonding with it. The
transport properties at equilibrium and non-equilibrium conditions 
were calculated using a Green's functions approach. The electronic states are
described by means of a semiempirical extended H\"uckel 
theory (EHT), where parameters are obtained from first principle 
calculations\cite{Cerda2000}. We did guarantee a very good agreement between
our electronic description and \textit{ab initio}  calculations, with errors less than 2\%
around the Fermi energy.\\
From the Hamiltonian of a molecule $H_{mol}$ and a contact $H_C$,
a coupled system (contact $C$ - molecule) can be described by
the composite Schrodinger equation 
\begin{equation}
\left[ (E+ i\eta)S - H \right]G(E) = 1, \quad  \eta \rightarrow 0^+
\end{equation}
where S is the normalization matrix and $G$ is the Green's function 
of the total system. By using the Green's function, the spectral function 
(density of states per unit energy) can be written as
\begin{equation}
A(E)= -\frac{2}{\pi}Im[G(E)]
\end{equation}
for the spin unpolarized case. Following this description, the local density of states
can be written as
\begin{equation}
\rho(r,E) = \sum_{\mu \nu} A_{\mu \nu}(E)\psi_{\mu}\psi_{\nu}
\end{equation}
where $\psi$s are the molecular wavefunctions, giving rise to the electronic density as
\begin{equation}
n(r) = \int_{-\infty}^{E_F}dE \rho(r,E)
\end{equation}
that considers the integration just until the Fermi energy level.

\begin{figure}[h!]
\centering
\includegraphics[scale=1.0]{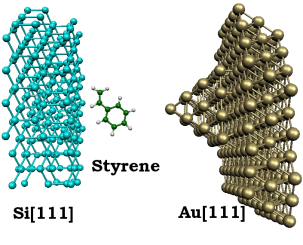} \\
\caption[System]{Atomic configuration of an 
organic molecule interacting, at a given optimal distance, with a semiconducting surface Si[111] and
a metallic electrode Au[111] used as a STM probe.}
\label{fig1}
\end{figure}

Now, if we want to study the consequences on
a specif\mbox{}ic part of the system due to the presence of the other parts,
the Green's function approach allows it by including the effect as perturbations
on the original Green function. 
This approach has been described by Williams \textit{et al}.~\cite{Williams1982} to study defects
in crystals as perturbations. They have showed that the problem
can be described by a Dyson equation and have obtained the Green's function of the
perturbed system. In our case, the Dyson equation leads to the perturbed
Green's function of the cluster due to the presence of the surface. 
The surface ef\mbox{}fects  are included as
\begin{equation}
G(E)=G^0(E)+G^0(E)\Sigma G(E)
\end{equation} 
where $G^0$ and $G$ are the Green's functions associated with the isolated
(unperturbed part)
and the supported cluster (perturbed) respectively, and $\Sigma$ is the self-energy matrix given by:
\begin{equation}
\Sigma =\tau G^0_s(E) \tau^{\dag}~.
\end{equation} 
The previous equation
 takes into account the ef\mbox{}fect of the surface, which could come as broadening and shifting of
 molecular energy levels (the size of $G$, $G^0$ and $\Sigma$ are def\mbox{}ined mainly from the
 molecular orbitals values). The term 
 $G^0_s(E)$ corresponds to
the Green's function of the isolated surface and $\tau$ is the coupling
matrix between the cluster and the surface. 
The non equilibrium regime, is considered when another electrode is added to
the system and an external voltage is applied between the surface and the electrode.
In this case, the representation for the supported cluster is changed by using the
correlation function $G^{<}=G\Sigma^{<} G^{<}$ to determine the charge density at a specif\mbox{}ic
bias $V$. The term $\Sigma^{<}=i(f(E-\mu_s)\Gamma^s+f(E-\mu_e)\Gamma^e)$ is
the lesser self-energy that depends on the Fermi function $f_i=f(E-\mu_i)$
of the contact $i$ (surface and electrode) for a specific chemical 
potential $\mu$ and the broadening matrix $\Gamma$
of the surface and the metallic electrode. 
The rate at which electrons goes from one contact to another through the
cluster is the transmittance $T(E,V)$. By considering only coherent transport
(no scattering), it is possible to show that $T(E,V)$ can be evaluated using the relation
\begin{equation}
 T\equiv Tr[\Gamma^sG\Gamma^eG^{\dag} ].
\end{equation}
Finally, the expression used to calculate the current through a system
like the one depicted in fig. \ref{fig1}, where a molecule, with discrete
energy levels, is located
between a semiconductor surface and an electrode (with a uniform distribution
of states), can be written as
\begin{equation}
I = \frac{2e}{h} \int_{-\infty}^{\infty} dE T(E)[f_1-f_2]~.
\end{equation}
The transmittance $T(E)$ is integrated in the energy interval associated
with each applied bias voltage and restricted by the 
Fermi function of the contacts.
For both cases 
(surface and electrode), local equilibrium is assumed and, in principle, 
the low temperatures approximation is used, so the integration can be 
performed between $\mu_1$ and $\mu_2$.\\
After the  calculation of the transmittance is performed, we did follow the treatment 
given by  Tian \textit{et al.}~\cite{Tian1998} to obtain
the conductance and the dif\mbox{}ferential conductance. At low temperature regime, 
the conductance will be proportional to the transmittance (at zero bias), where the
dif\mbox{}ferential conductance will be written as 
\begin{equation}
 \frac{dI}{dV}\approx G_0T(E_f+eV)
\end{equation}
where $G_0=2e^2/h$ is the maximum conductance constant. 
Here it is assumed that the electrostatic potential modif\mbox{}ies mostly the
chemical
potential of the metallic electrode. To describe the molecular electronic
properties we use the EHT approach, where the electronic Hamiltonian is parameterized
through the matrix elements $H_{\alpha i, \beta j}$, between
two atomic orbitals $\alpha$ and $\beta$ of atoms $i$ and $j$ respectively
and assumed to be proportional to the overlap $O_{\alpha i, \beta j}$
 and such that 
 $H_{\alpha i, \beta j} = K_{\alpha i, \beta j} \cdot O_{\alpha i, \beta j}$. 
According with Cerda \textit{et al.}~\cite{Cerda2000}, the parameter 
$ K_{\alpha i, \beta j}$ can be def\mbox{}ined from the on site energies using the form: 
$K_{\alpha i, \beta j}=K_{EHT} \cdot (E_{\alpha i} + E_{\beta j})/2$, where
$K_{EHT}$ is a f\mbox{}itting parameter usually obtained for solids and calculated
  from f\mbox{}irst principles.In our case this value was fixed at 1.75 for H, 2.8
 for C and 2.3 for Au and Si.\\   
In order to evaluate the accuracy of this model, we have compared the description
obtained with the EHT model, with a calculation based on a plane wave implementation
of the density functional theory with normalized pseudo potentials,
as implemented in CPMD~\cite{CPMD}. 
The exchange-correlation has been described by using the 
local spin density approximation (LSDA)~\cite{Lee1988,Becke1988}.
The atomic structure and energy minimization
of the isolated molecule and the
supported molecule in contact with the surface have been calculated
at the same theory level.

\section{Results}
In order to make a proper electronic characterization of the system,
we did start by describing every one of the
system components as isolated systems. The surface considered in this work is a 
semiconductor silicon with  a [111] orientation and no reconstruction.
Even though, it is experimentally more typical to find this surface
as reconstructed
in (2x1) and (7x7), depending on temperature
~\cite{Watanabe1998,Bolotov2001}, it has been also possible to obtain a Si[111]-
in a (1x1)
surface reconstruction \cite{Kahn1983}. 
For the unreconstructed surface, the dangling bonds are
associated to the discrete energy levels that can be observed at the band structure of
the Si[111] surface.
In the case of the molecule, we have selected styrene (\Styr~) as the organic
molecule in contact with the surface. This molecule has been previously
considered and experimentally studied as a possible
component in the design of new molecular devices~\cite{Guisinger2004}.
For the coupled case, the interaction between the organic molecule and the
Si[111] surface has been obtained from a geometrical
optimization from first principle calculations. The final atomic structure
feed to the transport calculation is the lowest energy geometry.
The structural optimization has been obtained considering the
local spin density approximation (LSDA) as the exchange and 
correlation terms ~\cite{Goedecker1996}. 
From the optimization, we have obtained two possible structural
configurations for the supported molecule, labeled as case A and B from now
and on. The optimization also defines the optimal distance 
between the molecule and the surface. The energy difference between both
configurations  is 15.8 eV, being case B the most stable. This result was calculated
using a computational package Gaussian 98 ~\cite{Gaussian98}

\subsection{\Styr~ on Si[111] :  (case A)} 
For the case named as A, the energy difference between the HOMO and LUMO
levels of  the free molecule is 0.95 eV. Whereas, in the coupled case,
the electronic gap is obviously reduced and
the optimal molecule-surface distance is found to be 2.5 angs.
By using the minimal structure, the electronic structure is calculated using the EHT
model (as previously discussed). A fair comparison with the electronic levels
from ab-initio calculations was found.
between them.
\begin{figure}[h!]
\centering
\includegraphics{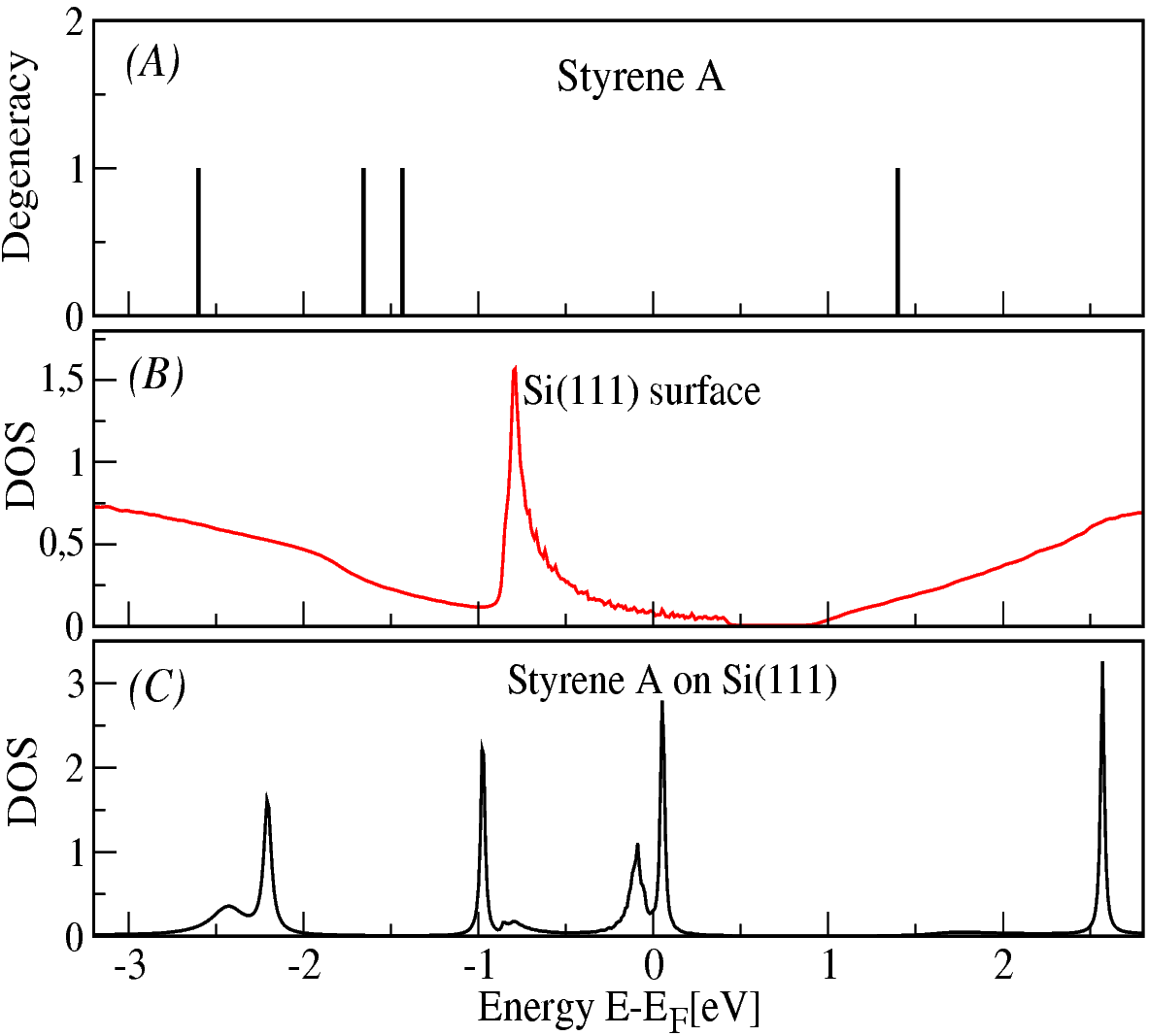}
\caption{Electronic structure of the \Styr~-Si[111] system.
Graph A shows the energy levels
of \Styr~ (A case) for the isolated case, whereas B shows the DOS
of Si[111] and finally C corresponds to the DOS of \Styr~  on ideal
Si[111].Energy values are shifted respect to the Fermi level.}
\label{fig2}
\end{figure}
In order to evaluate equilibrium properties we have considered
Figure~\ref{fig2}~that shows the electronic density of states of the
system calculated within the EHT approach. In Fig.~\ref{fig2}-A~ the 
electronic levels of the free molecule are presented. 
In Fig.~\ref{fig2}-B, the e-DOS of the Si[111] surface is shown. An 
electronic gap of 0.45 is
observed and dangling bonds are located around 0.7 eV below the Fermi
level. This description is in good agreement with results reported by 
Pandey~\cite{Pandey1976b} and Schl\"uter~\cite{Schluter1975},
obtained using a semiempirical tight-binding method and a self-consistent 
pseudopotential respectively. The presence of the dangling bonds is observed in
these cases with a shift of the states compared with our results and given
because of the parameterization of the considered model.
Finally, in Fig.~\ref{fig2}-C~ the e-DOS of the molecule interacting with
the surface is shown. The energy levels of the free cluster are broadened and 
red shifted (moved to the right side)
in the coupled  case 
As a consequence, there are more states close to
the Fermi level and the gap of the surface.\\ 
Once we have established the equilibrium conditions, 
a metallic electrode is included in the system to evaluate transport
properties. As a first approach,
an ideal Au(111) surface is considered as an electrode. Initially,
the electrode and surface are initially at local equilibrium and with chemical
potentials $\mu_e$=-5.26 eV and $\mu_s$=-4.8 eV
respectively~\cite{Allen1962}. 
\begin{figure}[h!]
\centering
\includegraphics[scale=2.0]{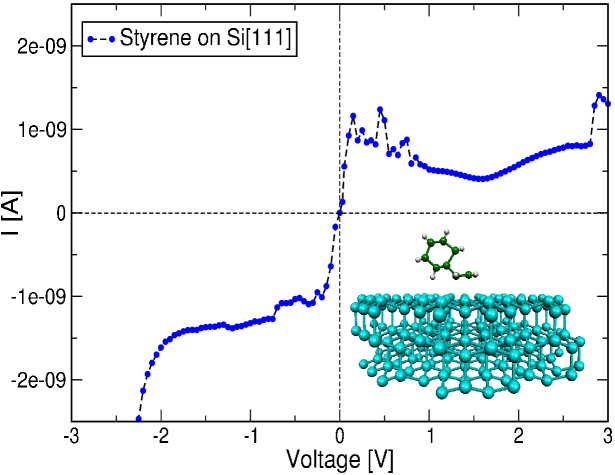}
\caption[total]{I-V characteristics of the system 
as a function of voltage and considering an interval of -3.0 and 3.0 V.}
\label{fig3}
\end{figure}
Under non-equilibrium conditions, an external
bias $V$ is imposed by changing the chemical potential of the metallic
electrode. Then the transmittance T(E,V) is calculated and the current can be
obtained integrating the transmittance in the energy range associated to each
bias. I-V curve is shown in Fig.~\ref{fig3} where the applied bias goes from -3 to 3
V. We can observe a NDR effect for a positive applied bias.
The intensity of the current starts to decrease at 0.7 V finding a
minimum around 1.5 V where the current starts to increase again. 
This effect has been also experimentally reported for styrene on 
Si[100]~~\cite{Guisinger2004}.
In this case, we have evaluated the states close to the Fermi level
finding a shifting effect (adjustment of the position of molecular levels) 
as a function of the applied bias in the 
DOS and a drop in the transmittance when the bias is larger than 0.5 V.
The molecular state, located next to the Fermi level (see Fig.~\ref{fig2}-C~), 
will be crucial when we have to explain the major
contribution to the NDR effect.
This state is the main peak in the DOS located in the
right side of the DOS respect to the Fermi level (that in this case was
considered at -4.8 eV for Si[111]) and the charge distribution associated to
it is located mainly between the molecule and the silicon surface.
In Fig.~\ref{fig4} the DOS of the styrene molecule is compared with the 
differential conductance dI/dV as a function of bias.    
\begin{figure}[h!]
\centering
\includegraphics[scale=2.0]{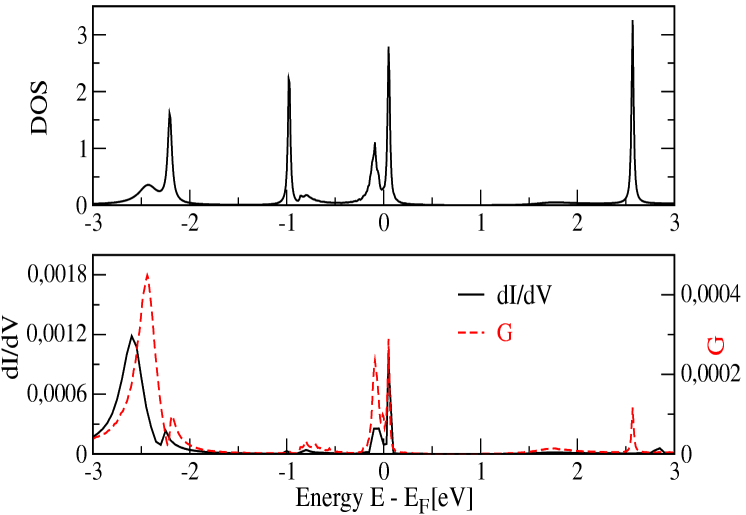}
\caption[total]{Calculated DOS and differential conductance dI/dV.}
\label{fig4}
\end{figure}
In the case
of dI/dV, we can  see not just the shift of the peaks respect to the e-DOS but
also the magnitude of some peaks clearly changing, when they are
compared to other peaks also observed in the e-DOS.\\
The dI/dV characteristic in Figure ~\ref{fig4} shows the molecular states 
that indeed contribute to the electronic transport in the system. 
In this context, the intensity of the two peaks located at +/- 0.2 eV 
respect to the Fermi level are clearly defined, in contrast to the state
located 1 eV below the Fermi level.  
Another parameter we did take into account is related to the doping on the surface.
To consider doped surfaces, a simple approximation  can be performed without the
inclusion of atoms in the system, such as P or S. This can be achieved by 
modifying the Fermi level of the
substrate. We have considered three cases where the chemical potential of the
surface is assumed to be $\mu_s=$-4.8 eV (initial case), -4.3 eV (n-type) and 
-5.3 eV (p-type). For each case, there are different equilibrium conditions
controlled by the chemical potential.
We have calculated the transmittance in the
system and then the current $I(V)$. These results have been depicted in
Fig.~\ref{fig5} where a voltage range from -3 to 3 V has been used. In this figure we can see
that the n-type surface shows a similar behavior compared to the non-doped
case ($\mu_s=-4.8 eV$), where the NDR effect at positive bias is observed.
In the case of the p-type
surface the result is rather different. The NDR is not found at positive bias and
even the behavior at negative $V$ is also different when it is compared with
 the previous cases. 
\begin{figure}[h!]
\centering
\includegraphics[scale=2.0]{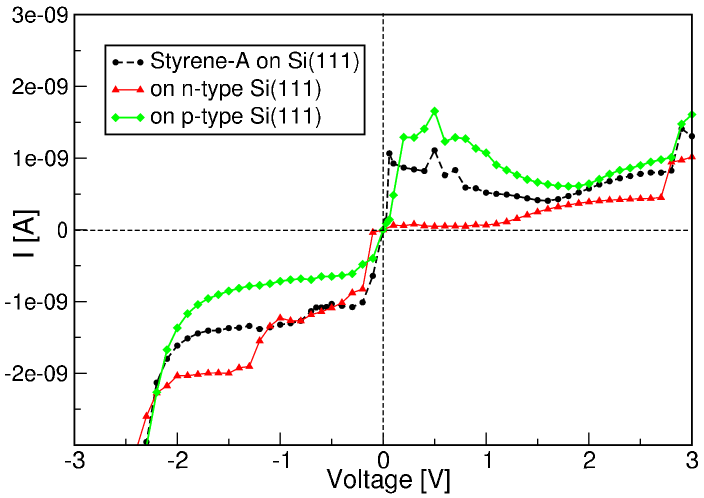}
\caption[total]{I-V characteristics for doped surfaces. To consider doped
  surfaces, the surface chemical potential $\mu_{s}$ is adjusted considering
  three cases: $\mu_s=$-4.8 eV (initial case), -4.3 eV (n-type) and -5.3 eV (p-type).}
\label{fig5}
\end{figure}
In order to understand the differences at each doped o undoped surface
Fig.~\ref{fig6}
shows the e-DOS and the differential conductance dI/dV.
Even considering that the peaks of e-DOS are clearly shown at each case, we can see
a displacement of the peaks as well as a
change in their magnitude. In the
case of dI/dV, the peaks are again shifted and changed in magnitude.
One of the important aspect of the doping effect, is that molecular states have been
also shifted according to each doped case and, in particular, the n-type surface 
($\mu_s=-4.3 eV$) have moved (red shifted) the
molecular states to the region of the calculated energy gap for the Si[111] surface
(see Fig.~\ref{fig2}).
\begin{figure}[h!]
\centering
\includegraphics[scale=2.0]{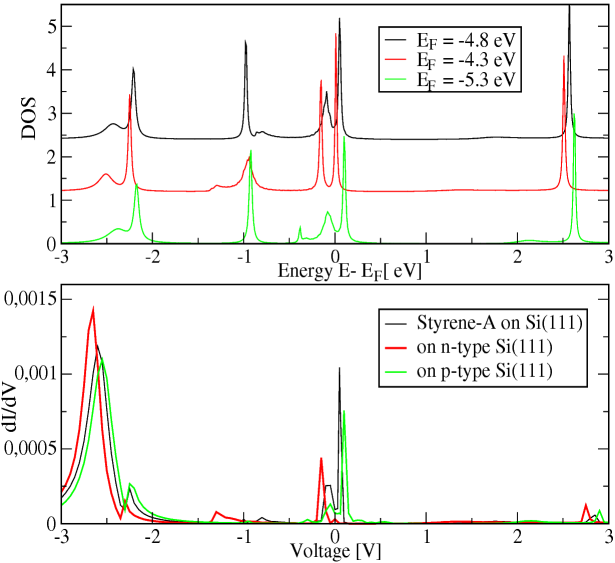}
\caption[total]{DOS and dI/dV characteristics for doped surfaces. To consider doped
  surfaces, the surface chemical potential $\mu_{s}$ is adjusted to consider
  initial case, a p-type and a n-type surface.}
\label{fig6}
\end{figure}
In order to differentiate the influence of the tip geometry, we did consider
two different geometries: a flat surface and a pyramidal structure.
The last geometry is shown in  Fig.~\ref{fig7} where the tip shape is
obtained with 30 Au atoms.
The molecule-tip distance has been fixed to 5.0
angs. For the flat electrode, we have considered a distance of 5.2 angs.
These distances are long enough to consider only tunneling current between 
the molecule and the tip/electrode.
\begin{figure}[h!]
\centering
\includegraphics[scale=2.0]{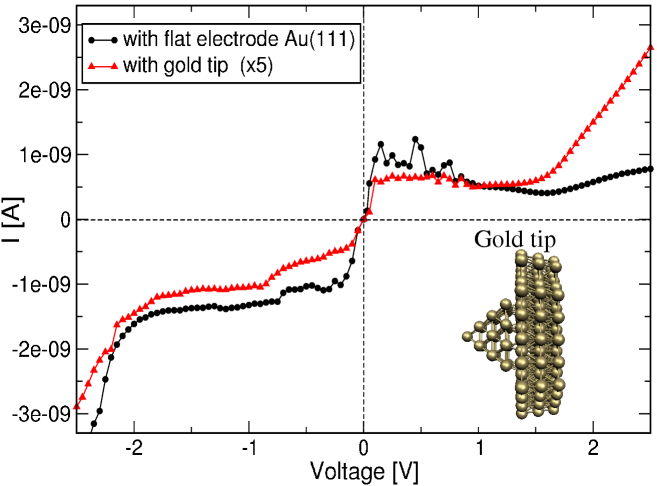}
\caption[total]{I-V characteristics considering different probes. Results
  using a flat electrode are compared two gold tips.}
\label{fig7}
\end{figure}
Fig.~\ref{fig7} shows the I-V characteristics with and without a tip. In the
case of the tip, the intensity of the current was multiplied by 5
for better comparison. 
We can see that at low voltages (positive and negative), the
behavior is rather similar between them, but this description changes when the bias is
higher in the positive direction. The NDR effect observed in the case of the flat
electrode, has been diminished until dissappear, showing in contrast
an almost constant behavior when a bias between 0.5 and 1.5 V is applied.
\begin{figure}[h!]
\centering
\includegraphics[scale=2.0]{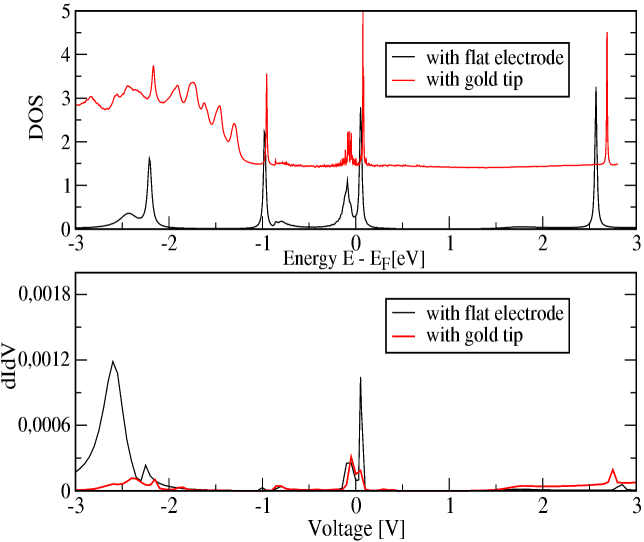}
\caption[total]{DOS and dI/dV characteristics considering different probes. Results
  using a flat electrode are compared two gold tips.}
\label{fig8}
\end{figure}
The calculated e-DOS and the dI/dV in Fig.~\ref{fig8} show the comparison
between the flat electrode and the tip,
in particular, a change in the magnitude of the peak located at 0.2 eV above the Fermi
level, where the intensity is diminished when the tip is considered.
There is also another difference mainly localized at lower
energies and can be attributed to the considered tip.\\

\subsection{\Styr~ on Si[111]: B case} 

The previous discussed molecular geometry is the one which has attracted
more attention when it is supported by Si[100] surface~\cite{Guisinger2004}.
Nevertheless, We have found a second,
stable configuration (B case) when the styrene molecule is interacting with the Si[111]
with no reconstruction.  
Is important to notice that the energetic
optimization shows that E(case B) $<$ E(case A), which means that this
configuration is much more stable. In this case, the optimal distance between the 
molecule (H atoms) and the surface (Si atoms) was found at 2.8 angs.
\begin{figure}[h!]
\centering
\includegraphics[scale=2.0]{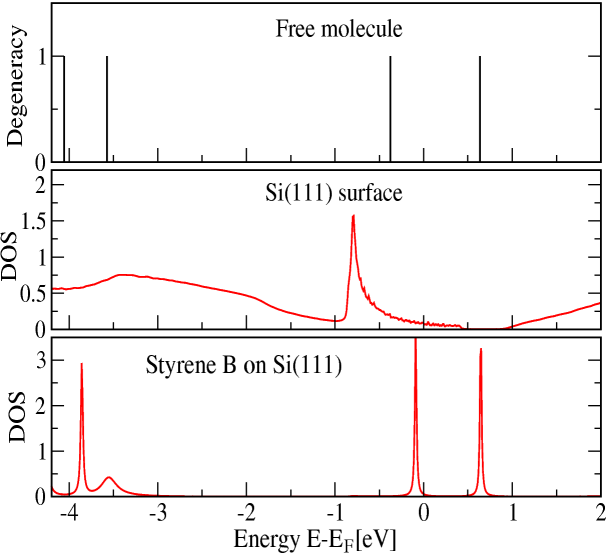}
\caption[total]{Electronic structure of the \Styr~(B case) -Si[111] system.
Graph A shows the energy levels
of \Styr~ for the isolated case, whereas B shows the DOS
of Si[111] and finally C corresponds to the DOS of \Styr~  on ideal
Si[111].}
\label{fig9}
\end{figure}
Figure~\ref{fig9} shows that energy levels associated with the case B free molecule
(graph A) and the e-DOS of the Si[111] surface (graph B) and the molecular
levels for the molecule interacting with the surface (graph C).
From the e-DOS of the supported molecule, we can see molecular levels
energetically located not only close, but also in the energy gap associated to the surface 
(graph B in Fig. ~\ref{fig9}).  The two molecular states localized around 0 and
0.7 correspond to the original LUMO and LUMO+1 states respectively and they play an
important role in the conduction phenomena.
After we have evaluated the electronic configuration at equilibrium conditions (without bias), 
the I-V curve is shown in Fig.~\ref{fig10} with the same voltage
range as case A. Interestingly, the I-V characteristic does not seem to show a similar
NDR effect when a positive bias is applied, like in the first case. Instead of
this effect, there is no an important contribution for the current at low
positive voltages until a bias of 2.0 V is applied. Instead, when the negative bias
is considered, there is current flow even at low voltages.
\begin{figure}[h!]
\centering
\includegraphics[scale=2.0]{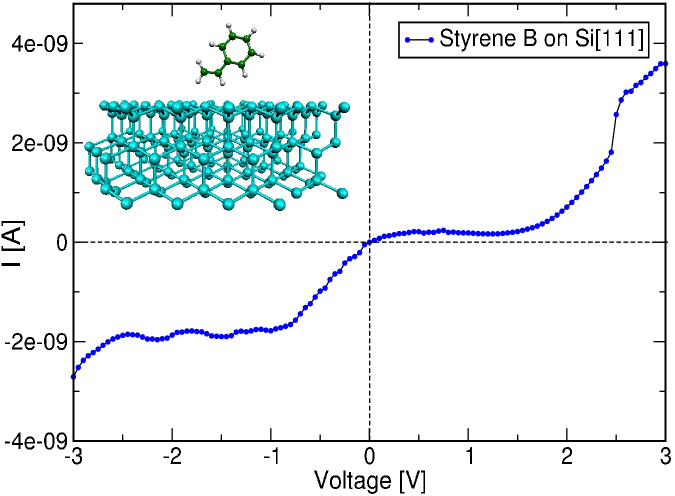}
\caption[total]{I-V characteristics of the system 
as a function of voltage.}
\label{fig10}
\end{figure}
If we want to know about the states that participate in the electronic
transport, it is possible to evaluate the differential conductance dI/dV 
and the conductance $G$ in Fig.~\ref{fig11}.
In this case, the contribution of the state located initially at
0.7 eV above the Fermi level, has a very small intensity in dI/dV characteristic
compared to other states. Even the conductance and the dI/dV of the state 
located at 0.2 eV  below the Fermi level, is also small in comparison to,
for example, the states located at 2.3 eV above the Fermi level. There is also
a spread distribution of the conductance and dI/dV characteristics 
for low negative values (between -1.0 and 0.2 eV below $E_F$). In this
case, there will be important to notice that this distribution is located in
the same energy interval
associated to the dangling bonds on the silicon surface. 
\begin{figure}[h!]
\centering
\includegraphics[scale=2.0]{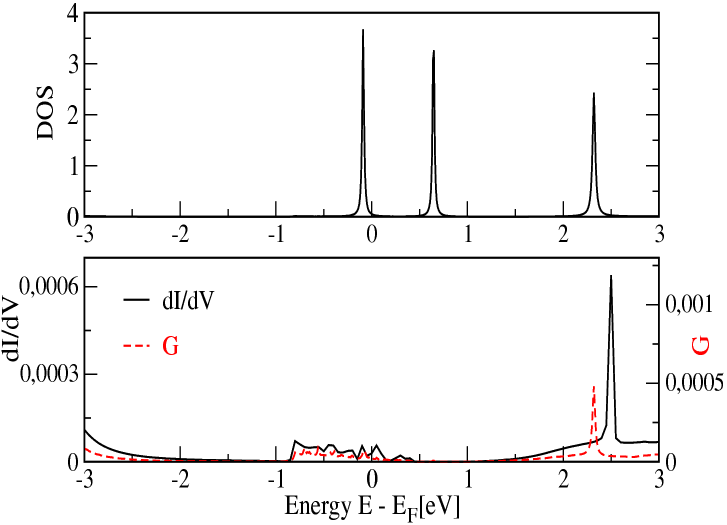}
\caption[total]{DOS and dI/dV characteristics of the system 
as a function of voltage.}
\label{fig11}
\end{figure}
Following the same lines than in the previous configuration, we did also study
the transport properties as function of surface doping.
As previously explained, we did obtain these conditions by 
modifying the chemical potential. Results
for the I-V characteristics are shown in Fig.~\ref{fig12}. This figure shows 
clearly different results
when the substrate is n-type or p-type doped. In the first case, we
have the results obtained for \Styr~ (B case) on Si[111] with a chemical
potential of -4.8 eV (original conditions). The current increases steadily
after an
applied bias of 2.0 V. When the surface is n-type doped (considering
$\mu_s= -4.3 eV$), there is no response in the current until 0.6 V, where
the intensity starts to increase rapidly with the positive applied bias.
A completely different behavior is obtained when the surface is p-type 
(which means that $\mu_s= -5.3 eV$ ). Here we can see that a NDR effect is
obtained for the current around  1-2 V, when even if the positive bias is
increasing, the current flow is decreasing in magnitude until a minimum value around 2 V. 
After this bias, the current increases its value again in a similar way like
in the two other doped surfaces. 
\begin{figure}[h!]
\centering
\includegraphics[scale=2.0]{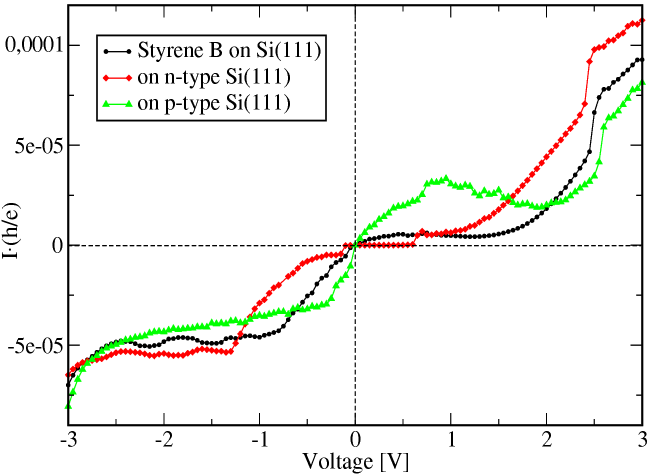}
\caption[total]{I-V characteristics for undoped (circles) and doped (diamonds
and triangles) surfaces. To consider doped
  surfaces, the surface chemical potential $\mu_{s}$ is adjusted to each case.}
\label{fig12}
\end{figure}
The figure ~\ref{fig13} shows the DOS and the dI/dV characteristic associated
with doped o undoped surface. Here, we notice that the energy has been adjusted to each
different Fermi level, so the zero energy position is indeed different for each condition. 
It is important to notice that even with a similar distribution of the
molecular states,
the differential conductance looks rather different. For the undoped surface,
there is no an important peak for dI/dV at low voltages and the main one is located at
2.4 V, with a spread contribution of states
 obtained around 0 V. If we consider the n-type Si[111] surface, 
we can detect an important peak located around 0 V within with the same 
spread distribution as for the undoped case. For the p-type
surface, we can see a small peak around 0.0 V and 2 important peaks, one located around
0.7 V and the other one at 2.5 V.      
\begin{figure}[h!]
\centering
\includegraphics[scale=2.0]{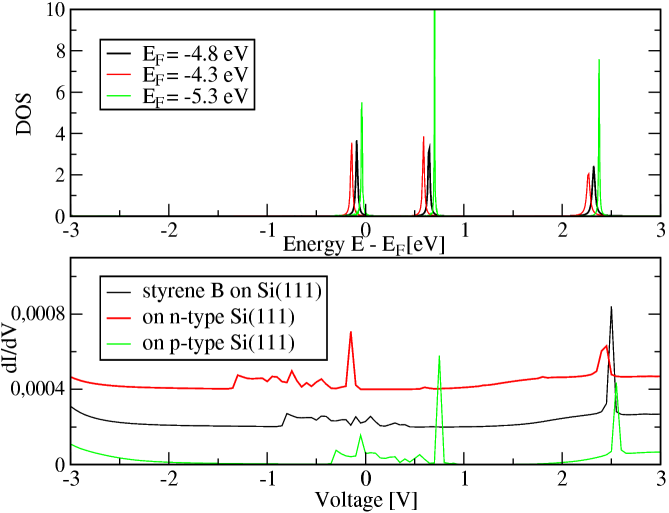}
\caption[total]{DOS and dI/dV characteristics for doped surfaces. To consider doped
  surfaces, the surface chemical potential $\mu_{s}$ is adjusted considering
  three cases. In the case of dI/dV, values were shifted in intensity for
  a better description}
\label{fig13}
\end{figure}
For all doped-undoped surfaces, the same peak, locate around 2.5, is clearly depicted
and is related to the state found at 2.4 eV above the Fermi level.
Again, the states related to the peaks close to the Fermi level, provide 
the most important contribution when a low voltage regime is considered. 

\section{Discussion and conclusions}
Here, we have reported a set of electronic transport calculations for a
semiconductor-cluster-electrode system. Initially we have obtained two
possible configurations (A and B cases) for the styrene molecule 
given by the minimization of the
energy associated to the surface-molecule system . The most stable configuration corresponds
to the B case and the A case corresponds to the atomic configuration of the
molecule used for Si[111] surface, as it is shown in Fig.~\ref{fig3}.
For the A case, we have obtained a NDR effect when a
positive bias is applied. This can be understood based on the shifting of the 
molecular levels, close to the Fermi level, when the voltage is applied.
This effect is modulated when the surface has been doped, which changes the chemical
potential. This effect is more important
for the case of a p-type surface, as it is observed in Fig.~\ref{fig3}, when
compared to the n-type surface, where the effect 
is eliminated from the I-V characteristic. This difference is
achieved because of the shift of the molecular levels according to each doped surface.
 For the n-type surface (with $\mu_s$=-4.8 eV), the electronic molecular levels are
 energetically located inside the energy gap of the silicon surface, so
there is no channel available for the electronic transport at low positive bias
(see dI/dV curve for n-type surface in Fig.~\ref{fig6}). Another result,
shows the influence of the tip in the electronic transport, where if the tip
is changed, the NDR effect could be diminished with respect to the flat electrode.
In order to compare both conditions, we have used a similar
molecule-tip/electrode distance (5 angs.).
This can be understood from the potential created by the tip-surface, which is
smaller than the potential created by the electrode-surface case. A smaller
electric field created for a sharp tip promotes a very small shift of the
energy levels of the molecule and then it is more difficult to move the
states to the energy interval associated to the energy gap of Si[111].
For the second configuration (B case), the situation is rather different
 because there is no NDR when the undoped surface is considered. This can be concluded
by observing that the main peak occurs at 0.7 eV above $E_F$ and therefore
can not be used as an electronic channel due to the energy gap of the Si[111].
 Also, to control the NDR effect, we have considered doped surfaces,
where it is possible to detect the NDR effect for the p-type surface but not for
the n-type case. This difference is explained because the number of the channels
available, and close to the Fermi level, is lesser for the undoped surface than for
the p-type surface. Then the shifting of the 
states produced by
the applied bias diminishes the contribution of the peak that is energetically
displaced to the energy gap of the doped surface. 
In Fig.~\ref{fig13}, we can see that the undoped surface
and the n-type case do not show important peaks close to the Fermi level
in the dI/dV characteristics and the most important contribution comes from
the state located 2.3 eV above $E_F$. If we consider the p-type surface,
we can see not only the contribution given by the states related to the
dangling bonds of the surface around the Fermi level, but also one peak
located at 0.7 eV above $E_F$ and another one at 2.5 eV above $E_F$.\\

In summary, the extended H$\ddot{\textrm{u}}$ckel model, used to describe the 
electronic properties in the system, it is in good agreement when compared with
{\it ab initio} calculations. We have found two stable configurations for the
styrene molecule in contact with the surface and both of them can provide 
interesting properties like the NDR effect. This effect is modulated
not only by the electronic molecular levels but also their positions respect to the
energy gap of the silicon surface. 
A semiconductor surface has interesting properties respect to metallic ones,
 allowing to control transport with the appropriate conditions like doping. In our case,
we have a NDR effect of the system at low positive voltage. This effect is
also more important when a strong electric field has been considered to
shift the molecular states (flat electrode produces a stronger electric field
than a sharp tip). The comparison between flat and sharp electrodes has been
performed for both configurations finding the same NDR suppression when a sharp
metallic tip is considered.   
These results strength the idea to use organic molecules as possible
components to storage information in electronic devices.


M. E. G. acknowledges the kind hospitality at the 
IPICYT. This work has been  supported by the DFG through the SPP1153. 
A.H.R. also thanks the support of Conacyt Mexico under the grant J-46247-F.
S. E. B. acknowledges the scholarship given by IPICyT and the University of Kassel.

\end{document}